\documentclass[fleqn,10pt]{wlscirep}
\usepackage[utf8]{inputenc}
\usepackage[T1]{fontenc}
\usepackage{graphicx}
\usepackage{latexsym}
\usepackage{amsmath}
\usepackage{amssymb}
\usepackage{amsfonts}
\usepackage{color}
\usepackage{bm}
\usepackage{verbatim}
\usepackage{ulem,cancel}

\definecolor{MS-color2}{RGB}{128,0,128}

\title{Flux flow spin Hall effect in type-II superconductors with spin-splitting field }

\author[1,2]{Artjom Vargunin}
\author[1,*]{Mikhail Silaev}
\affil[1]{Department of
Physics and Nanoscience Center, University of Jyv\"askyl\"a, P.O.
Box 35 (YFL), FI-40014 University of Jyv\"askyl\"a, Finland}
\affil[2]{Institute of Physics, University of Tartu, Tartu, EE-50411, Estonia}

\affil[*]{mikesilaev@gmail.com}

\begin{abstract}
 We predict the very large spin Hall effect in type-II superconductors which mechanism is drastically different from the 
previously known ones.  We find that 
in the flux-flow regime the spin is transported by the spin-polarized Abrikosov vortices moving under the action of the Lorenz force in the direction perpendicular to the applied electric current. Due to the large vortex velocities the spin Hall angle can be of the order 
of unity in realistic systems based on the high-field superconductors or the recently developed superconductor/ferromagnetic insulator proximity structures. 
 We propose the realization of high-frequency  pure spin current generator based on the periodic structure 
of moving vortex lattices. We find the patterns of  charge imbalance and spin accumulation generated by moving vortices, which can be used for the electrical detection of individual vortex motion.  The new mechanism of inverse flux-flow spin Hall effect is found based on the 
{ driving force acting on the vortices in the presence of injected spin current which results in the generation of transverse voltage.}
\end{abstract}
\begin{document}

\flushbottom
\maketitle
%
%
\thispagestyle{empty}


\section*{Introduction}
 The spin Hall effect (SHE) is currently one of the basic tools in spintronics used for the 
 generation and detection of pure spin currents \cite{Sinova2015}. 
 Although it has quite a rich variety of applications, from the fundamental point of view
 there has been only two known mechanisms leading to the spin Hall effect: 
 {\bf (i)} the spin-orbital interaction in semiconductors and heavy metals and
 {\bf (ii)} the Zeeman spin splitting in graphene close to the neutrality point making the electrons and holes to carry different spin polarizations \cite{Abanin2011,Abanin2011a,Wei2016}. 
 Here we suggest the third fundamental mechanism combining the  specific properties of the electronic spectrum
 in superconductors with spin-splitting field and the coherent dynamics of the superconducting order parameter manifested through the flux flow of Abrikosov vortices under the action of the external transport current.  

  The non-equilibrium properties of superconductors with spin-splitting fields have become a hot topic in the field of superconductivity\cite{SilaevRMP2017}.
  Such systems are characterized by the spin-dependent electron-hole asymmetry of   Bogolubov quasiparticles\cite{Tedrow1971}. 
  Recently it has been realized that this feature allows for the generation of spin   accumulation\cite{Silaev2015,Bobkova2015,Krishtop2015,Virtanen2016,PhysRevB.98.024516,1706.08245}, which is 
  robust against the usual spin-flip and spin-orbital scattering relaxations. 
  This mechanism explains many experimental observations of long-range non-local spin signals in 
  mesoscopic superconducting wires generated by the injected current from the ferromagnetic or 
  even non-ferromagnetic electrodes\cite{Huebler2012,Wolf2014,Quay2013,Quay2016}. 
  In this paper we demonstrate the possibility of not only the long-range spin accumulation but also the non-decaying pure spin current generation using the properties of  superconductors with spin-splitting fields. 
  
 In principle, the paramagnetic spin-splitting of Bogolubov quasiparticles appears inevitably due to the Zeeman effect in any superconductor subject to the magnetic field\cite{Maki1966,Huebler2012,Quay2013}. However,
 the magnetic field simultaneously leads to the orbital effect, inducing the center-of mass motion of the Cooper pairs due to the Meissner effect.
  %
  The relative magnitude of the paramagnetic shift and the orbital kinetic energy of the Cooper pair 
  is determined by the parameter introduced by Maki \cite{Maki1966} (referred later as the Maki parameter)
  $\alpha_0 = \mu_Bc/(eD)$, where $\mu_B$ is the Bohr magneton, $D$ is the diffusion coefficient, $e$ 
  is the electron charge and $c$ is the light velocity.
  Usually the orbital effect in superconductors dominates over the paramagnetic one, provided that the second critical 
  field $H_{c2}$ is not too high so that $\mu_B H_{c2}\ll k_BT_c$. In this case the Maki parameter is small $\alpha_0\ll 1$. Exceptions are the high-field superconductors 
  were the Zeeman shift can become relatively large
  at fields not exceeding $H_{c2}$
  \cite{Chandrasekhar1962, Clogston1962, Maki1966,saint1969type,Gurevich2010, Cho2011}.
The paramagnetic effect can be   significantly enhanced due to the 
  geometrical confinement in thin superconducting films 
  \cite{PhysRevLett.24.1004,Huebler2012,Quay2013,Kolenda2016}. 
     Alternatively, the spin splitting in superconductors can be induced by the exchange interaction of conduction electrons with localized magnetic moments, e.g. aligned magnetic impurities\cite{fulde_ferrell.1964}. 
    Recently, the  systems consisting 
  of  superconducting films grown on the surfaces 
  of ferromagnetic insulators like EuS \cite{Wolf2014,Kolenda2017,Strambini2017} and GdN \cite{PhysRevB.97.224414}
    have been fabricated. 
    There exchange field $\bm h_{eff}$ in the superconducting film is induced due to the scattering of conductivity electrons from the 
    ferromagnetic insulator interface\cite{PhysRevB.38.8823}. 
  Such systems are currently studied quite actively as the possible platforms for the 
  advanced radiation sensing technology\cite{1706.08245, 1709.08856} and 
  quantum computing with Majorana states\cite{PhysRevB.98.020501}.   
     
\begin{figure}[!htb]
 \centerline{\includegraphics[width=0.8\linewidth]{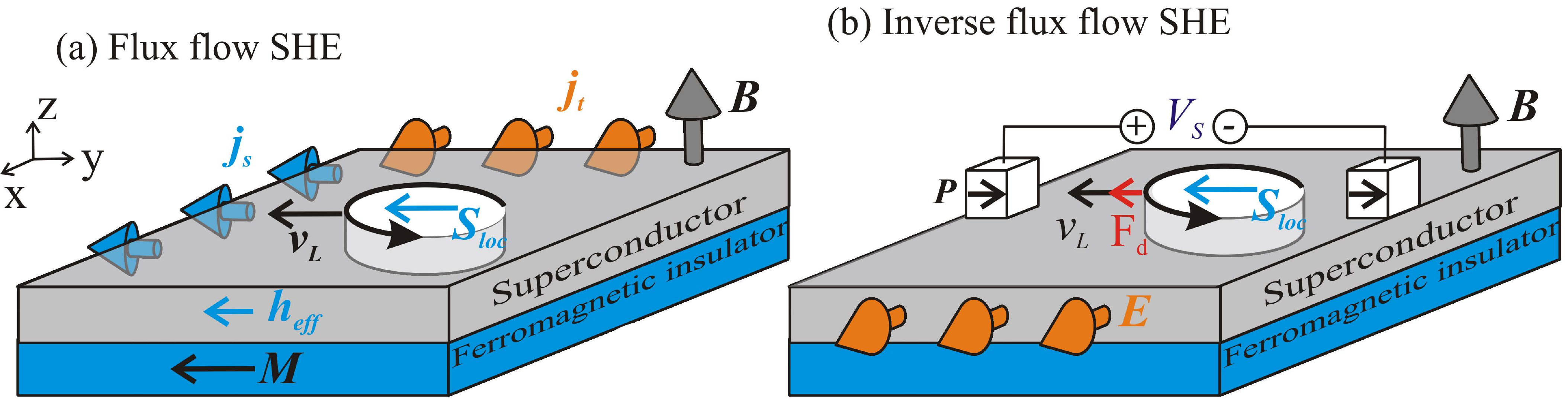}}
 \caption{\label{Fig:Cartoon} 
 {\bf The mechanisms of direct and inverse  flux-flow spin Hall effects. }
The magnetic field $\bm B$ perpendicular to film plane creates Abrikosov vortices in superconductor shown by white cylinders surrounded by the circulating current. Vortex cores contain localized normal phase which acquires spin 
polarization ${\bm S}_{loc}$ due to the splitting field induced  by ferromagnetic insulator and Zeeman effect.  
(a) Direct flux-flow SHE. The interaction of transport $\bm j_t$ and superconducting current circulating around vortex generates the Lorentz force 
driving vortex lattice motion in the transverse direction with the velocity $\bm v_{L} \perp \bm j_t$. Vortex motion results in the transverse spin current $\bm j_s$.
(b) Inverse flux-flow SHE. The spin-dependent bias $V_s$ can be generated by biasing the ferromagnetic  electrodes with polarization $\bm P$, attached to the  superconductor. The induced spin accumulation gradient $\nabla \mu_z$  produces the driving force on the vortex lines $\bm F_d\parallel \nabla \mu_z$. In result vortex lattice tend to move in the direction $\bm v_{L}\parallel \nabla\mu_z$ and produce the average electric field in transverse direction $\bm E = \bm B\times \bm v_{L}/c$. 
   }
 \end{figure}

  The most well known paramagnetic effects in spin-singlet superconductors are the 
  first-order transition into the normal state\cite{Chandrasekhar:1962,PhysRevLett.9.266} and the second-order transition into the 
  inhomogeneous superconducting state induced by the spin-splitting field $\bm h_{eff}$. 
 The inhomogeneous state (FFLO) 
   suggested by  Fulde, Ferrell\cite{fulde_ferrell.1964} and Larkin, Ovchinnikov\cite{larkin_ovchinnikov.1964}
   is realized in the narrow window of parameters and suppressed by impurities\cite{Aslamazov} which hinders its experimental realizations\cite{Buzdin2005}. However the  first-order transition into the normal state driven by the Zeeman splitting has been detected in thin aluminum films\cite{PhysRevLett.24.1004}. 
   In this paper we focus on the more robust nonequilibrium phenomena which generically appear in the presence of any spin-splitting field in the spin-singlet superconductor\cite{1706.08245} .   
  In particular, we consider the film of type-II superconductor which can host Abrikosov vortices. The example of such setup setup is shown schematically in Fig. \ref{Fig:Cartoon}. It consists of the thin superconducting film deposited on the magnetic insulator which creates spin splitting of the conduction electron subbands in the superconductor due to the effective exchange interaction $\bm h_{eff}$. In addition there is a 
  magnetic field $\bm B$ 
  directed perpendicular to the film plane to create vortices. The total spin splitting field is given by the superposition  $\bm h=\mu_B \bm B + \bm h_{eff}$, so the single-particle Hamiltonian becomes 
{  $H = (i\hbar \nabla + e \bm A/c)^2/(2m) + \hat{\bm \sigma}\bm h$, where $ \bm A$ is the vector potential} and 
  $\hat{\bm \sigma}$ is the vector of spin Pauli matrices.
  Superconductors with the total spin splitting field $\bm h = \bm h_{eff} +\mu_B \bm B $ coming both due to Zeeman shift and internal exchange  are characterized  by  the renormalized Maki parameter 
  $\alpha = \alpha_0 h/(\mu_B H_{c2} )$.
  It can become large  $\alpha\sim 1$ if the total spin splitting is close to the paramagnetic 
  depairing threshold $h\sim k_B T_c$. Such strong spin splitting has been recently obtained in superconductor/ferromagnetic insulator 
  proximity structures used for the generation of the long-range spin accumulation in the non-local spin valve
  geometries\cite{Wolf2014,Kolenda2017,Strambini2017, 1706.08245, 1710.10833}.  
 Due to the large exchange field this regime can be achieved even if the Zeeman effect is 
  small, that is when $\mu_B B\ll k_B T_c$. 

  Although we focus on the bilayer system, 
  the regime when $\alpha\sim 1$ is also possible in high-field bulk superconductors 
   where the spin splitting comes solely from the Zeeman effect\cite{Maki1966,saint1969type}. 
Similar {behavior} can be observed in magnetic superconductors\cite{Buzdin1984}, 
such as borocarbides \cite{Mueller2001} where weak ferromagnetic ordering is possible \cite{Bluhm2006} 
and vortex cores can host localized paramagnetic moments \cite{DeBeer-Schmitt2007}. 
  In this systems the internal exchange field plays the same role as the proximity-{induced} 
  spin splitting in the bilayer system and weak pinning facilitates flux-flow regime.

  %

     Below we demonstrate that $\alpha$ becomes the only relevant parameter 
  which determines the amplitude of the pure spin current generated by 
  the vortex motion. 
  { The latter can be characterized by} the spin Hall angle $\theta_{sH} = ej_s/j$, where 
  $j_s$ is the induced  spin current and  and $\bm j= \bm j_t$ is the 
  charge current equal to the transport current generated by the external source. The spin Hall angle can be estimated as $\theta_{sH}\sim \alpha $.
  At the paramagnetic threshold  $h\sim k_BT_c$ it can reach {$\theta_{sH}\sim 1$}
  which is much larger than the record values $\theta_{sH}< 0.1$ obtained in the heavy metal 
  spin current generators \cite{Sinova2015}. 
  
  The above result is rather surprising because the maximal spin splitting $h\sim k_BT_c$ 
  is very small as compared to the Fermi energy $E_F$, 
  since in usual superconductors $E_F/k_BT_c \sim 10^2 - 10^3$. In this case
  the polarization, which is the relative difference between spin-up/down conductivities 
  is rather small $ \sim h/E_F\ll 1$. This limit yields vanishing  
  spin-polarized component  of the resistive current. 
      However, { it is the vortex motion which} generates much larger spin current in the transverse direction  
  $\bm j_s \perp \bm  j $ as  explained below. 

   The scheme of the flux-flow direct  
   spin Hall effect is shown in  
    Fig. 	\ref{Fig:Cartoon}a. 
    Here, we assume that the superconductor with spin-splitting field and vortices  is subject to the
    transport charge current $\bm j$ generated by the external source. This transport current induces the Lorenz force acting on the vortex lines in the direction perpendicular to current $\bm F_L\propto \bm j\times \bm B$.
    Provided that the Lorenz force  overcomes the pinning barrier, vortices start to move in the transverse direction
 with the velocity $\bm v_L\perp \bm j$. 
Taking into account the spin polarization $S_{loc}$ which exists inside each vortex core due to paramagnetic response, this motion generates the transverse pure spin current 
 $\bm j_s \approx n_v  \bm v_L S_{loc}$,  where $n_v = B/\phi_0$ is the vortex density, $\phi_0$ is flux quantum. 

 Vortex cores in diffusive superconductors can be though of as the normal metal tubes, 
 of the diameter determined by the coherence length $ \xi$. 
 In the presence of spin splitting field, the vortex cores contain localized spin 
 $S_{loc} \sim \chi_n h \xi^2$ per unit vortex length, 
 where $\chi_n= N_0$ is the normal metal paramagnetic susceptibility and 
 $N_0$ is the Fermi-level density of states.  
To estimate $j_s$ we substitute the flux-flow vortex velocity $v_L= - cE/B $
and get {$\theta_{sH}  \sim ch/(eDH_{c2})\sim\alpha$,
so that $\alpha$} appears to be the only small parameter 
limiting the spin current generation. 
The physical reason for large $\theta_{sH}$ lies in the fast motion of vortices which can be compared e.g. with 
the Drude-model electron drift velocity $\bar v = \sigma_n E/ne$, where
the conductivity is $\sigma_n = e^2N_0 D$. At $B\approx H_{c2}$ we have the relation 
$ v_L \approx (E_F/k_BT_c) \bar v \gg \bar v$. Therefore spin polarization can be transported much faster 
by moving vortices than by electrons drifting along the electric field.  

Along with the direct SHE we propose also the scheme of the inverse flux-flow SHE shown in Fig. \ref{Fig:Cartoon}b. 
 The mechanism is based on the injection of spin-polarized quasiparticle current into the superconductor by applying the voltage through the spin-filtering ferromagnetic electrodes with polarization $\bm P$. The resulting spin-dependent voltage  $V_s$ generates the spatially-inhomogeneous non-equilibrium spin accumulation  which we hereafter denote $\mu_z$.  Its gradient  $\nabla \mu_z$ will be shown to produce the longitudinal force acting on the spin-polarized vortex cores pushing them towards one of the ferromagnetic electrodes. The vortex lattice motion with velocity $\bm v_L$ generates electric field in the transverse direction $\bm E \parallel \bm B\times \nabla\mu_s$ thus providing the novel mechanism of inverse SHE.

\section*{Model}
To quantify  effects discussed above we use the framework of Keldysh-Usadel 
 theory \cite{Schmid1975,Belzig1999} describing the spin current and spin accumulation induced by the vortex motion in the 
  usual $s$-wave spin-singlet superconductor in the diffusive regime\cite{1706.08245}.
 We consider the range of magnetic fields close to $H_{c2}$, neglecting screening and using the Abrikosov solution for the moving  vortex lattice. 
We will show that in addition to the large average spin current 
there is also the oscillating part 
which can be considered as the high-frequency source of the spin current 
at the nearly-terahertz range  \cite{Embon2017}.


 We use the formalism of quasiclassical Green's functions (GF)
 \cite{Schmid1975,Belzig1999} generalized to describe the non-equilibrium spin states in diffusive superconductors \cite{1706.08245},
 $\check{g} = \left(%
 \begin{array}{cc}
 \hat g^R &  \hat g^K \\
 0 & \hat g^A \\
 \end{array} 
 \right)$, where $\hat g^{R/A/K}$ are the retarded/advanced/Keldysh components which are 
 the matrices in spin-Nambu space and depend on two times and a single spatial 
 coordinate variable $\check{g} = \check{g} (t_1,t_2,{\bm r})$. 
 We consider general expressions for the spin density deviation from the normal state one $S$ and  the spin current density $\bm j_s$ projected on the spin-splitting field direction $\bm h$. 
 These quantities are given by the following general expressions 
  \begin{align} \label{Eq:Spin}
 &  S (t,\bm r) = -(\pi\chi_n/8) {\rm Tr} [ \hat\tau_3\hat\sigma_h \hat g^K (t,t,\bm r) ]
   \\ \label{Eq:SpinCurrent}
 &  \bm j_s (t,\bm r)  = (\pi\sigma_n/8 e^2) {\rm Tr}[\hat\sigma_h( \hat g \circ  \hat \partial_{\bm r} \hat g )^K](t,t,\bm r) .
  \end{align}
 Hereafter, $\hat\sigma_h =( \hat \sigma_1h_x + \hat \sigma_2h_y + \hat\sigma_3 h_z)/h$ is the operator of the spin projection on the  spin-splitting field direction,  $\hat\sigma_i$, $\hat\tau_i$ with $i=0,1,2,3$ are the Pauli matrices in spin and Nambu spaces, 
  the symbolic time-convolution operator is given by $ (A\circ B) (t_1,t_2) = \int dt A(t_1,t)B(t,t_2)$,  
  the covariant differential superoperator is defined by 
  $\hat \partial_{\bm r}= \nabla  -  i \tilde e[\hat\tau_3{\bm A},]_t$,
  where $\tilde e = e/\hbar c$
  and the two-time commutator is defined as $[X, g]_t= X(t_1) g(t_1,t_2)- g(t_1,t_2) X(t_2)$, 
  similarly for anticommutator $\{,\}_t$.

   The general expression for $\bm j_s$ can be simplified using the following steps. 
  First, due to the normalization condition $(\check g\circ \check g) (t_1,t_2)= \check I \delta (t_1 - t_2)$, where $\check I$ is the unit matrix in Keldysh-Nambu-spin space,
  we introduce the parametrization of Keldysh component in terms of the distribution function 
  $\hat g^K = \hat g^R\circ \hat f- \hat f\circ \hat g^A$. 
  Second, we introduce mixed representation
  $\check{g} (t_1,t_2) = \int_{-\infty}^{\infty} \check{g}(\varepsilon, t) 
  e^{-i\varepsilon (t_1-t_2) } d\varepsilon/2\pi$, where $t=(t_1+t_2)/2$ and use gradient expansion of the time convolution product in Eq. (\ref{Eq:SpinCurrent})  as explained below. 

  In the flux-flow regime we assume that vortices move with the constant velocity $\bm v_L$. In the zero-order approximation the distribution function is equilibrium 
  $\hat f(\varepsilon) = f_0(\varepsilon)\hat\tau_0\equiv \tanh[\varepsilon/(2k_BT)]\hat\tau_0 $. Similarly, the spectral functions have their equilibrium forms in the 
  frame moving together with vortices 
  $\hat g^{R/A} (\bm r) \approx \hat g^{R/A}_0(\bm r - \bm v_L t) $.
 This approximation yields zero spin current which is absent in equilibrium spin-singlet superconductors.  
  Thus, we 
  need to consider  corrections in the linear-response regime which is realized provided the vortex velocity $\bm v_L$ is small enough to neglect Joule heating, pair breaking or vortex-core shrinking effects
  \cite{nonlin1,nonlin2}.
   For this purpose we take into account first-order terms in the gradient expansion of time convolutions 
  \cite{Larkin1986,KopninBook} as well as the non-equilibrium corrections to the spectral functions 
  $\hat g^{R/A}_{ne}$ and the distribution function
  $\hat f_{ne} = \hat f - f_0 \hat\tau_0$. In result we get the two parts of spin current $\bm j_s = \bm j_{s1} + \bm j_{s2} $ given by 
   \begin{align}\label{Ifullmixed}
  &\bm j_{s1,2}=  \frac{\sigma_n}{16e^2}
  \int_{-\infty}^{\infty}{\rm Tr} 
  [\hat\sigma_h  \hat{\bm J}^K_{1,2}] d\varepsilon,
  \\ \label{Eq:JK1}
    &\hat {\bm J}^K_1 = f_0(\hat {\bm J}^R_{ne} - \hat {\bm J}^A_{ne}) ,
    \\ \label{JT}
       &\hat{\bm J}^{R/A}_{ne} = (\hat g^{R/A} \hat\partial_r \hat g^{R/A})_{ne}   
       - 
   \frac{i\hbar}{2} (\partial_t\hat g^{R/A}_0\partial_\varepsilon\hat\nabla\hat g^{R/A}_0 - 
   \partial_\varepsilon\hat g^{R/A}_0\partial_t\hat\nabla\hat g^{R/A}_0),
      \\ \label{Eq:JK2}
  &\hat {\bm J}^K_2 = 
     \hat\partial_r\hat f - 
  \hat g^R_0\hat\partial_r\hat f\hat g^A_0 
  +
  \hat {\bm J}^R_0\hat f_{ne} - \hat f_{ne}\hat {\bm J}^A_0.
   \end{align}
  Although expressions (\ref{Ifullmixed},\ref{Eq:JK1},\ref{JT},\ref{Eq:JK2}) look quite involved, 
  different contributions to the current there have clear physical meanings. 
  The first part of the spin current $\bm j_{s1}$ is determined by the non-equilibrium 
  corrections to the spectral quantities  while it contains only the equilibrium  distribution function.
  Here 
  $\hat{\bm J}_{ne}^{R/A}$ given by Eq.(\ref{JT}) are the deviations of the spectral current densities  from equilibrium.
  In the charge sector these corrections 
   yield the Caroli-Maki part of the flux-flow conductivity \cite{PhysRev.164.591}. The  important difference is that the charge current 
   is determined by the corrections induced by the order parameter distortions in the  moving vortex lattice while they do not contribute to the spin current. 

  The 
  first term in the r.h.s. of (\ref{JT})  incorporates corrections to the spectral GF $g^{R/A}_{ne}$ as well as the electric field term which appears from the expansion of the covariant differential operator\cite{Silaev2016a} $\hat \partial_r \hat X=\hat\nabla \hat X  + 
    e {\bm E} \{ \hat\tau_3, \partial_\varepsilon \hat X \}/2 $ and $\hat\nabla X = \nabla \hat X - i\tilde e {\bm A} [\hat\tau_3, \hat X]$. 
 Here we 
  use the gauge with zero electric potential such that electric field is given by $\bm E = - \partial_t \bm A/c$. 
  The second term in the r.h.s. of (\ref{JT})
comes from the linear-order expansion of time convolution. It  contains the equilibrium spectral GF in the moving frame $\hat g^{R/A}_0 (\bm r -\bm v_L t) $.  
       The second part of the spin current 
  $\bm j_{s2}$
    is determined by the non-equilibrium 
  distribution function. This contribution is analogous to the Tompson term in the flux-flow conductivity\cite{PhysRevB.1.327}.
  Importantly, in the general case the differential operator in the r.h.s. of (\ref{Eq:JK2}) contains correction from time convolution expansion so that $\hat\partial_r \hat f= \nabla \hat f + e \bm E \hat\tau_3\partial_\varepsilon f_0$. 
  This correction gives contribution to the charge current but drops from the expression for the spin current (\ref{Ifullmixed}).
  The nonequilibrium GF is determined by the Keldysh-Usadel equation\cite{Schmid1975,Belzig1999} which should be solved together with the self-consistency equations. In general this problem is very complicated and has never been approached even numerically. However, the regime of 
 high magnetic fields {$H_{c2}-B \ll H_{c2}$} allows for significant simplifications based on the existence of the  Abrikosov vortex lattice solution for the superconducting order parameter. In this case it is possible to find analytically 
 nonequilibrium corrections to the spectral functions $\hat g^{R/A}$ and the components of distribution function $\hat f$. 

 First of all, we employ the analytical expression for the order parameter distribution in the moving vortex lattice.
 Assuming the particular directions of vortex velocity $\bm v_L =  v_L\bm y$ and electric field $\bm E = E \bm x$ we choose the time-dependent vector
 potential in the form  ${\bm A} =  B x{\bm y} -  cE t{\bm x} $. 
   Then the order parameter is given by superposition
 of the first Landau-level nuclei $\mathcal{L}(x) = \exp (- x^2/2L_H^2)  $, so that 
 $\Delta= b_0e^{-2i \tilde e E tx}\sum_n C_n e^{inp (y-v_Lt)} \mathcal{L}(x- nx_0)$. 
 Here { $b_0$ is field dependent amplitude of the gap, $x_0=p L_H^2$ determines the 
 distance between neighbour superconducting nuclei} 
 and  $L_H=1/\sqrt{2\tilde eH_{c2}}$ is the magnetic length. For the triangular lattice 
 { $C_{n+1}= e^{i(-1)^n\pi/4}$}, $pL_H=\sqrt{\pi\sqrt{3}}$ and for 
 the square one $C_{n}= 1$, $pL_H=\sqrt{2\pi}$.

\begin{figure}[!t]
 \centerline{\includegraphics[width=0.7\linewidth]{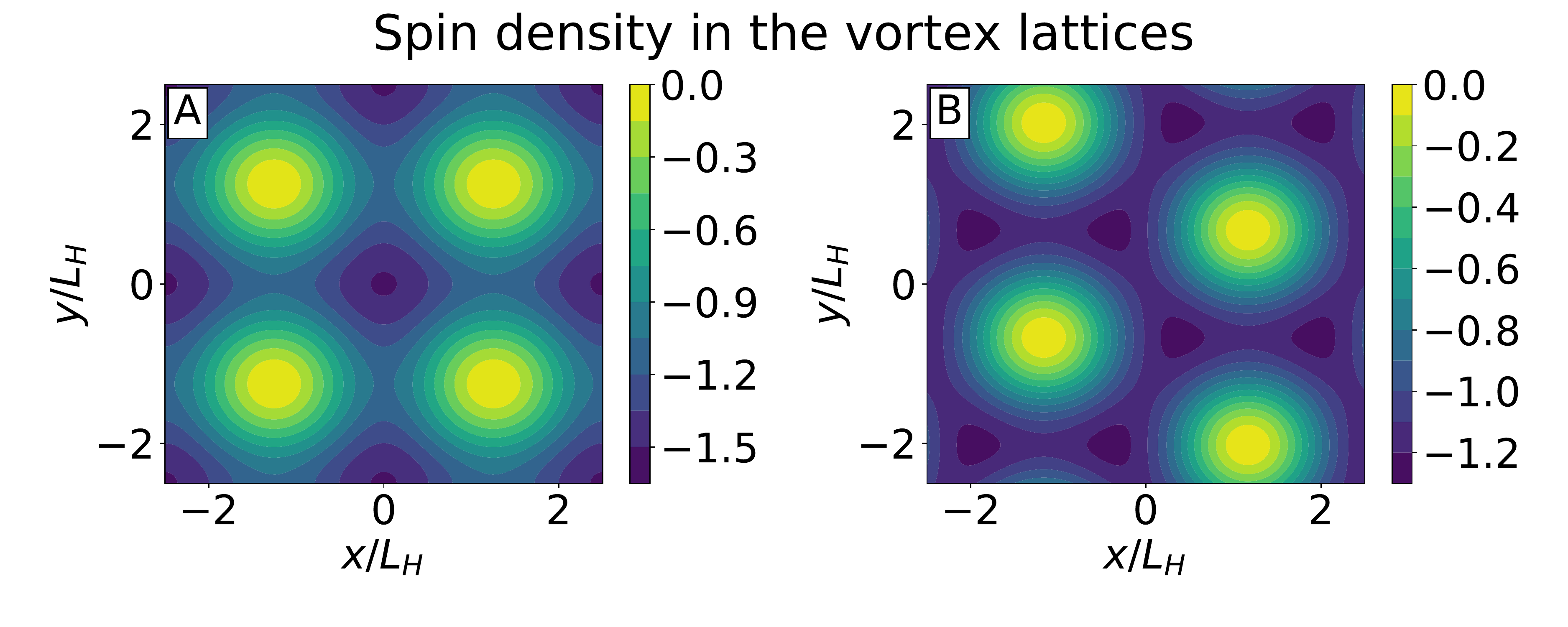} }
 \caption{ \label{f2a} 
 {\bf Spin density modulation in the vortex cores }.  
 Normalized deviation of the total spin density from the normal metal background, $S/S_n$, on square (A) and triangular (B) lattices. Here $S_n=-\chi_n h$ is spin polarization of the normal metal and $S/S_n$ 
 is shown in the units of dimensionless order parameter amplitude $\langle\Delta^2 \rangle/(k_BT_c)^2$.
 Calculations were performed at low-temperatures, $T\ll T_{c}$, for effective Maki parameter $\alpha=0.5$.
    }
 \end{figure}

Second, we use the known solutions for the 
equilibrium spectral functions in the vortex lattice near the upper critical field\cite{Silaev2016a}. Here we take into account the spin-splitting field by shifting the
quasiparticle energies according to $\varepsilon_{\sigma} = \varepsilon - \sigma h$, where $\sigma=\pm $. Then the 
 spin-up $g^{R/A}_{0+}$ and spin-down $\hat g^{R/A}_{0-}$ GFs are given by
  \begin{equation}\label{Eq:SpectralFunctions}
  \hat g^R_{0\sigma}(\bm r,\varepsilon) = \left[ 1 + \frac{|\Delta|^2}{2(iq+\varepsilon_\sigma)^2}\right] \hat\tau_3 + 
  \frac{i|\Delta|\hat\tau_2e^{-i\varphi\hat\tau_3 }}{iq + \varepsilon_\sigma},
  \end{equation}  
 and 
 $\hat g_0^A = -\hat\tau_3\hat g_0^{R+}\hat\tau_3$ for the advanced GF. 
 Here $q = \tilde e {\hbar}H_{c2} D$ and the order parameter is 
 $\Delta = |\Delta| e^{i\varphi} $. 
 These spin-polarized spectral functions provide the description 
 of equilibrium spin density modulation in a superconductor with spin-splitting field 
 in the presence of vortex lattices. The periodic spin density patterns calculated for the typical cases of triangular and square lattices are shown in the Fig. \ref{f2a}.  The spin polarization  
demonstrates enhancement at the vortex cores and suppression between vortices where the order parameter is larger. Thus even in the regime of dense vortex lattices there is an excess spin polarization $S_{loc}$ localized in the vortex cores. It is natural to expect that the motion of such spin-polarized vortices will produce pure spin currents.
 %
  Below we demonstrate the presence of these spin currents by an explicit calculation in the flux-flow regime considering the non-equilibrium situation when the vortex lattice moves under the action of the 
  transport current $\bm j_t$. We will calculate the spin current density induced by the vortex motion as well as the non-equilibrium spin accumulation and charge imbalance near the vortex cores. 
  
 \section*{Results}
 \subsection*{Spin current}
  
   To find the contribution $\bm j_{s1}$ to spin current density using Eq.(\ref{Eq:JK1}) we need the non-equilibrium corrections for the spectral functions. In the linear response approximation, assuming that the non-equilibrium corrections are small we can find them  
 using the normalization conditions for quasiclassical propagators $(\hat g^R \circ \hat g^R) (t_1,t_2)= \hat I \delta (t_1 - t_2) $. The calculation detailed in the   Supplementary Information yields the expression for the first term in Eq. (\ref{JT}) through the derivatives of the  equilibrium GFs 
 $(\hat g^R\hat\nabla\hat g^R)_{ne} = i\hbar \hat\nabla
 ( \partial_t \hat g^R_0\partial_\varepsilon \hat g^R_0  - 
 \partial_\varepsilon \hat g^R_0\partial_t \hat g^R_0  )/4$.
 Hence the spectral spin current density is given by 
 $\hat{\bm J}^R = i\hbar ( \partial_\varepsilon \hat  
 g_{0}^R \hat\nabla\partial_t \hat g_{0}^R    - 
 \partial_t\hat g_{0}^R \hat\nabla\partial_\varepsilon 
 \hat g_{0}^R )/2$.
  In the linear response approximation we keep only the first-order time derivatives 
  $\partial_t \hat{g}^{R/A}_0 = - (\bm v_L \nabla) \hat{g}^{R/A}_0$.
 In the considered high-field regime $H_{c2}-B\ll H_{c2}$  
  the order parameter is described by the Abrikosov vortex lattice solution. Thus we can use spectral functions $\hat g^{R/A}_0 = \hat g^{R/A}_0(\bm r- \bm v_L t, \varepsilon)$  given by the Eq.(\ref{Eq:SpectralFunctions}). 
 Substituting the above spectral current density into the Eq.(\ref{Ifullmixed}) and transforming the energy integral
 to the summation over Matsubara frequencies we obtain the first part of the spin current 
 \begin{equation}
 \bm j_{s1} = -\frac{\hbar \sigma_n}{16e^2} 
  \frac{{\rm Im}\Psi^{(2)}}{(\pi k_BT)^2}
  {\rm Re} [ \Delta (\hat\Pi \partial_t\Delta)^\ast - 
  \partial_t\Delta(\hat\Pi\Delta)^\ast ],
  \end{equation}
 where $\Psi = \Psi [1/2+ (q+ih)/(2\pi k_BT) ] $ is digamma 
 function, $\Psi^{(n)}(z)= d^{n} \Psi(z)/dz^n$ and 
 $\hat \Pi = \nabla -2i\tilde e\bm A$. 
 This part of the spin current has the non-zero space- and time-average $\langle \bm j_s \rangle = \langle \bm j_{s1} \rangle$  determined by the following expression which derivation is shown in the  Supplementary Information, 
 \begin{equation}\label{meanjs}
 \langle \bm j_s \rangle = - \bm v_L \sigma_n  
 \frac{\hbar \langle\Delta^2\rangle{\rm Im} \Psi^{(2)}}{(4\pi k_BT  e L_H)^2 }, 
 \end{equation}
where {$\langle\Delta^2\rangle = \sqrt{\pi}|b_0|^2 L_H/x_0 $} is the order parameter average over the vortex lattice cell. 
 At the same time the average spin density deviation from the normal state induced by the superconducting correlations 
 reads as
 $\langle S \rangle = - \chi_n \langle\Delta^2\rangle  {\rm Im} \Psi^{(1)}/(4\pi T)$. 
 Therefore, at low temperatures $T\to 0$ we obtain the asymptotic relation 
 $\langle \bm j_s \rangle = - 2\bm v_L \langle S \rangle /(1+\alpha^2) $.

 The spin current magnitude  depends on the order parameter amplitude $b_0$ which is determined by the magnetic field. 
 In the limit of large Ginzburg-Landau parameter we get the usual expression for the average gap
  function\cite{Caroli1966, Silaev2016} 
  $\langle\Delta^2\rangle = -8\pi k_B T q \delta B{\rm Re} \Psi^{(1)} /(\beta_L H_{c2}{\rm Re} \Psi^{(2)}) $, 
 where $\delta B = H_{c2}-B$ is the deviation of external field from the upper critical one,
  the Abrikosov parameter equals $\beta_L=1.16$ for the triangular and 
$\beta_L =1.18$ for the square lattice\cite{Kleiner1964}. 
 In the limit of low temperatures
 it can be simplified to $\langle\Delta^2\rangle = (1-B/H_{c2})(4q^2/\beta_L) (1+\alpha^2)/(1-\alpha^2)$
 yielding the following analytical expression for the spin Hall angle, { $\theta_{sH} = e\langle \bm j_s \rangle /\bm j_t$, as a function of the average magnetic induction at low temperatures,}
  \begin{equation} \label{Eq:SpinHallAngle}
  \theta_{sH}(B) = - \frac{4\alpha}{\beta_L(1-\alpha^4)}\left(1 - \frac{B}{H_{c2}} \right).
  \end{equation}
  The growth of $\theta_{sH}(B)$ with decreasing $B$ given by Eq.(\ref{Eq:SpinHallAngle}) 
   close to $H_{c2}$ should continue at lower fields until the order parameter between vortices 
   becomes fully developed at  $B\approx 0.3 H_{c2}$. In this regime we expect 
   {$\theta_{sH}\propto \alpha$} without any small parameters so that $\theta_{sH}\sim 1$ for large 
   exchange splitting $h\sim k_BT_c$. At smaller fields $B\ll H_{c2}$ the spin Hall angle should decrease as $\theta_{sH} \propto B/H_{c2}$, being proportional to the concentration of vortices. Besides that, according to Eq. (\ref{Eq:SpinHallAngle}) large spin Hall angle can be obtained already in the regime $(1-B/H_{c2})\ll 1$ provided that $1-\alpha\ll 1-B/H_{c2}$. Note that we restrict our consideration to $\alpha<1$ when the superconducting transition at $B=H_{c2}$ is of the second order   \cite{Maki1969,saint1969type}.      

  Now let us consider the second contribution to the spin current $\bm j_{s2}$ which according to the Eq.(\ref{Eq:JK2}) 
is determined by the non-equilibrium components of the distribution function generated by the vortex lattice motion. 
Due to the smallness of the order parameter in the regime 
$H_{c2}-B \ll H_{c2}$ this spin current contribution can be written in terms of the non-equilibrium spin accumulation 
 $\bm j_{s2} = (\sigma_n/e^2) \nabla \mu_{s1}$.
 Here $\mu_{s1} = \int_{-\infty}^{\infty} f_{T3}(\varepsilon) d\varepsilon/2$ is the contribution to the spin accumulation determined by spin-dependent 
component of the distribution function $f_{T3} = {\rm Tr} [\hat\sigma_h \hat f]/4$, which
 can be considered as the spin-dependent shift of the chemical potential \cite{SilaevRMP2017}.
{  Since $\mu_{s1}$ has to be periodic function, its gradients cannot provide 
 non-zero space-average or time-average spin current, however this contribution is also of interest since it produces AC component of $\bm j_s$. To find it, we need to solve the kinetic 
 equation for the spin imbalance, which is similar to 
 that for the longitudinal distribution function 
 \cite{Silaev2016a}
{
 \begin{equation} \label{KineticEqT31}
 \nabla^2 f_{T3} = -  e \bm E\bm J_{se} \partial_\varepsilon f_0
 -  \frac{1}{8D} \partial_\varepsilon f_0 {\rm Tr}[\hat\sigma_h\partial_t\hat \Delta(\hat g^R_{0} - \hat g^A_{0} )].
 \end{equation}
Here $\bm J_{se} = 
 {\rm Tr} [\hat\tau_3\hat\sigma_h
 (\hat{ \bm J}^R_0- \hat{ \bm J}^A_0) ]/8$ is spectral spin-energy current density
 \cite{PhysRevB.98.024516} and the gap operator in the second term of the r.h.s. is 
 $\hat \Delta = i|\Delta|\hat\tau_2 e^{-i\varphi\hat\tau_3}$. 
For the general-form Abrikosov vortex lattice, Eq.(\ref{KineticEqT31}) can be solved analytically yielding
\begin{align} \label{}
 & \mu_{s1} = \tilde e\hbar v_LH_{c2}F_0\sum_n\sin(py + \phi_n)F (x- nx_0) 
 \\
 & F(x) = \int_{-\infty}^\infty e^{-p|x-x^\prime|}
 \mathcal{L}(x^\prime - x_0)  \mathcal{L}(x^\prime ) dx^\prime,
 \\
 & F_0 = \langle\Delta^2\rangle 
 \frac{pL_H}{2\sqrt{\pi}} 
 \left( \frac{{\rm Im}\Psi^{(2)}}{(2\pi k_B T)^2}
  - \frac{{\rm Im}\Psi^{(1)}}{2\pi qk_B T} \right)
 \end{align}
and $\phi_n = {\rm arg}(C_n^\ast C_{n+1})$, see Supplementary Information. 

 \begin{figure}[!t]
 \centerline{\includegraphics[width=0.7\linewidth]{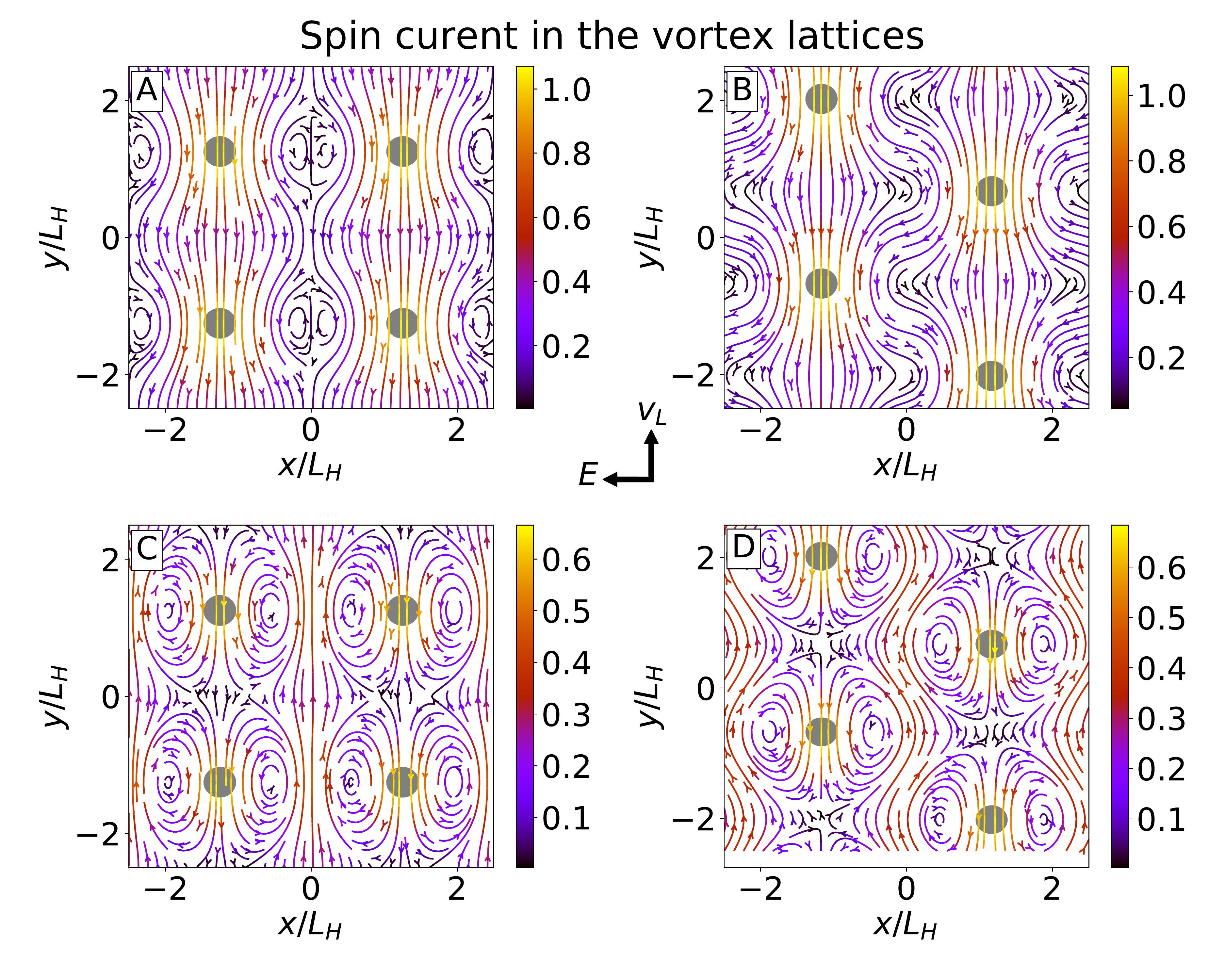} }
  \caption{  \label{f2b} 
   {\bf Spin current density generated by the vortex lattice motion.}  
  (A,B) The total spin current $\bm j_s$ for square and triangular spin lattices generated by the vortex lattice motion, normalized by  $v_L \chi_n \langle \Delta^2\rangle/(k_BT_{c})$. 
(C,D) Deviation of the net spin current from its spatial
  average $\tilde{\bm j}_s=\bm j_s - \langle \bm j_s \rangle$. Gray circles correspond to the position of the vortexes. 
 Left/right columns describe the case of the square/triangular lattices, respectively. 
 Arrows between panels indicate the direction of the vortex velocity $\bm v_L$ and average electric field $\bm E$. 
 Calculations were performed at low-temperatures, $T\ll T_{c}$, for $\alpha=0.5$. 
   }
 \end{figure}

The overall distributions spin currents are shown  in  
Fig. \ref{f2b} produced using the 
Matplotlib package\cite{Hunter2007} for two different vortex lattice geometries. 
 Here one can see that the spin current mostly flows along the vortex chains with  maximal current concentrated in the vortex cores.
This result confirms our initial qualitative picture shown in Fig. \ref{Fig:Cartoon} 
that the spin is transported by the moving spin-polarized vortex cores. 
In addition, in Figs. \ref{f2b}C,D one can see a non-trivial distribution of the 
spatially-periodic part of the current $\tilde{\bm j}_s = \bm j_s - \langle \bm j_s\rangle$,
 which is  important for the AC spin current generation discussed below.
The periodic part $\tilde{\bm j}_s$
 forms two standing eddies localized close to the vortex core similar to that which are formed by the
 low-Reynolds  viscous flow past a cylinder.

{\subsection*{Spin accumulation and charge imbalance}

 \begin{figure}[!t]
 \centerline{\includegraphics[width=0.7\linewidth]{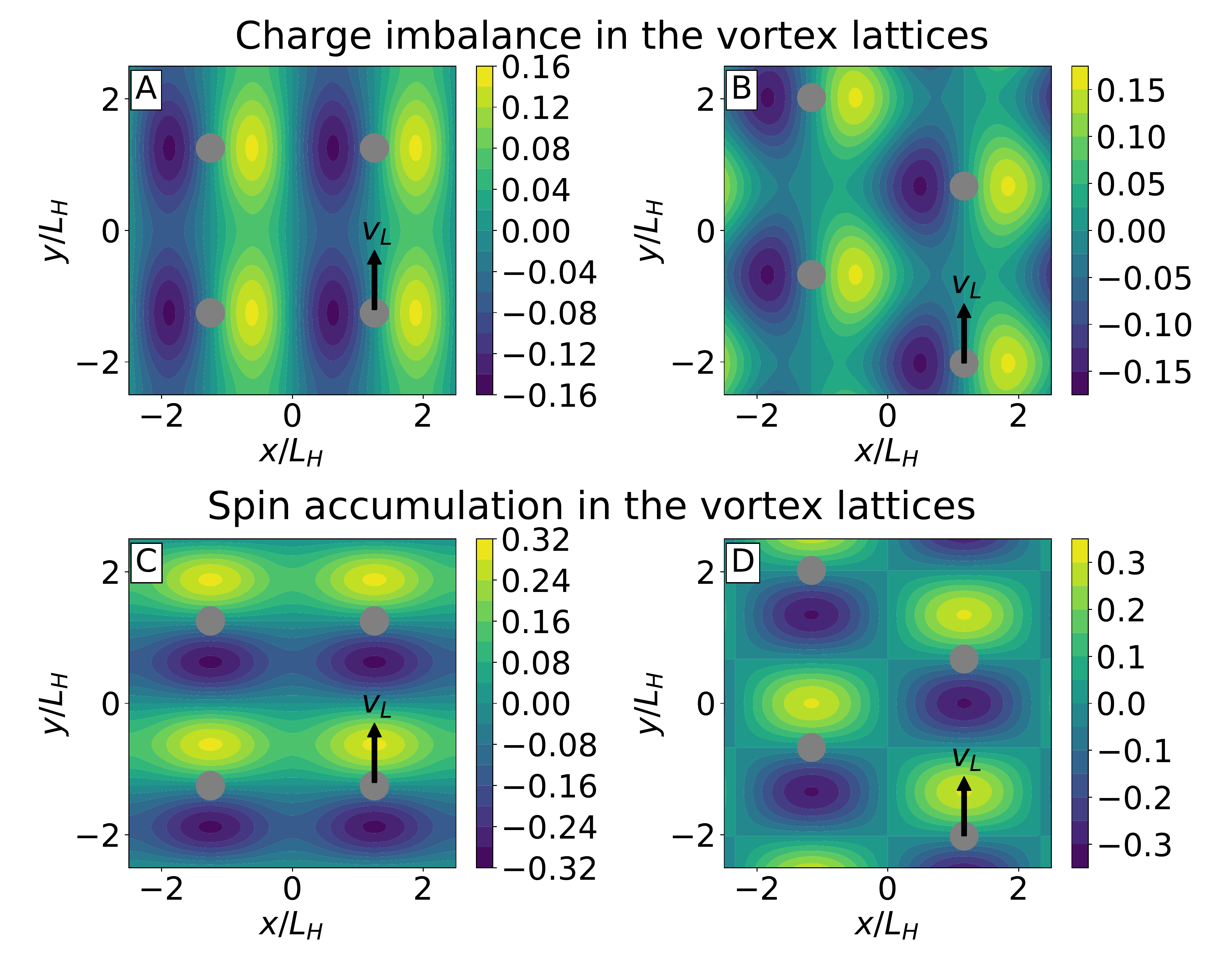}
 }
 \caption{ \label{f3} 
   Charge imbalance $\mu$ (A,B) and spin accumulation $\mu_z$ (C,D) generated by the moving vortex lattices. Both quantities are normalized to 
 $\hbar v_L\langle\Delta^2\rangle /(L_H q^2)$. Left/right columns describe the case of the square/triangular lattices, respectively.       
 Gray circles correspond to the position of the vortex cores and black arrows indicate the direction of the vortex velocity $\bm v_L$. 
  Calculations were performed at low-temperatures, $T\ll T_{c}$, for $\alpha=0.5$.
  } 
  \end{figure}

Besides generating the spin current, moving vortices produce other types of non-equilibrium states in the superconductor, such as the charge imbalance and the non-equilibrium spin accumulation which we denote as $\mu$ and $\mu_z$, respectively. These quantities has been widely used as the experimentally observable characteristics of the non-equilibrium superconducting states both in spin -degenerate\cite{yagi2006charge, hubler2010charge, Tinkham1972,
Tinkham1972a,clarke79,clarke80,PhysRevLett.28.1363, PhysRevB.12.4909}
and spin-split systems \cite{Quay2013,quay2015-qsr,
Quay2016, Huebler2012, Kolenda2016,Kolenda2017,
Wolf2014}. 
General expressions for charge imbalance and spin accumulation in terms of the quasiclassical GF read as
  \begin{align} \label{Eq:mu}
  & \mu=-(\pi /8){\rm Tr}\hat g^K_{ne}(t,t,\bm r)
  \\ \label{Eq:muZ}
  & \mu_s=-(\pi /8){\rm Tr}[\hat\tau_3\hat\sigma_h \hat g^K_{ne}](t,t,\bm r)
  \end{align}
  where $\hat g^K_{ne}$ is  non-equilibrium part of Keldysh GF. 
 To find $\mu$ and $\mu_s$ using the expressions (\ref{Eq:mu},\ref{Eq:muZ}) we employ the mixed representation with the first-order gradient expansion of the non-equilibrium part of  Keldysh GF, $\hat g^K_{ne}=(\hat g^R\hat f-\hat f\hat g^A)_{ne}-i\hbar\partial_\varepsilon f_0\partial_t(\hat g^R_0+\hat g^A_0)/2$. The spin accumulation $\mu_z$ is determined by spin imbalance mode $f_{T3}$ and time derivative of spectral GF so that close to $H_{c2}$ we have
\begin{align}\label{muz}
&\mu_s=-\mu_{s1}+\frac{\hbar v_L}{4}\frac{{\rm Im}\Psi^{(2)}}{(2\pi T)^2}\partial_y|\Delta|^2.
\end{align}
To obtain the second term in (\ref{muz}) we integrated $\partial_\varepsilon f_0{\rm Tr}[\hat \tau_3\hat\sigma_3\partial_t(\hat g^R_0+\hat g^A_0)]$ over energy by parts and transformed result to the sum over Matsubara frequencies. The details of this calculation can be checked in the Supplementary Information.  

To calculate charge imbalance generated by  moving vortex lattice we notice that term with time derivatives of spectral GF in $\hat g^K_{ne}$ is traced out, while non-equilibrium corrections to spectral functions are of importance. The latter can be found with the help of normalization condition which results in the leading order in $|\Delta|$ in ${\rm Tr}\hat g^{R/A}_{ne}=\pm\frac{i\hbar}{4}{\rm Tr}([\hat\tau_3,\partial_t \hat g_0^{R/A}]\partial_\varepsilon \hat g_0^{R/A} )$. Therefore 
\begin{align}
&\mu=-\mu_1-\frac{\hbar v_L}{8}\frac{{\rm Re}\Psi^{(2)}}{(2\pi T)^2}\partial_x|\Delta|^2,
\end{align}
where $\mu_1=\int_{-\infty}^\infty f_Td\varepsilon/2$ is contribution to the electrostatic potential from transverse component of distribution function, $f_{T} = {\rm Tr} [\hat\tau_3 \hat f]/4$. For high magnetic fields  kinetic equation for  $f_T$ reads as  
\begin{align} \label{KineticEqT}
  \nabla^2 f_T=-e\partial_\varepsilon f_0\nabla( {\cal D}_T{\bm E}) + \frac{1}{8D} \partial_\varepsilon f_0 {\rm Tr}[\hat\tau_3\partial_t\hat \Delta(\hat g^R+\hat g^A)] ,
  \end{align}
where ${\cal D}_T= {\rm Tr}( \hat\tau_0\hat\sigma_0 - \hat\tau_3\hat g^R\hat\tau_3\hat g^A)/8$. By using Abrikosov vortex-lattice solution, spatially periodic solution of kinetic equation can be found analytically, the reader can consult Supplementary Information for the details of this derivation. 

Distributions of $\mu$ and $\mu_z$  generated by the moving square and triangular vortex lattices are shown in  Fig. \ref{f3}. 
 The patterns of charge imbalance agree with the qualitative picture  suggested by Bardeen and Stephen \cite{PhysRev.140.A1197}
 where the vortex motion is accompanied by the generation of dipolar-like electric field near the vortex core, corresponding to the electric dipole directed perpendicular to the vortex velocity $\bm v_L$. 
 On the contrary, the "spin dipoles" corresponding to the patterns of $\mu_z$ are directed along $\bm v_L$. 
 Note also, that spin accumulation is proportional to the generalized Maki parameter, while $\mu$ remains finite when $\alpha\to 0$, that is  paramagnetic effects are neglected.

\subsection*{Inverse flux-flow spin Hall effect}

 We suggest the new mechanism of the inverse flux-flow SHE which is based on the previously unknown effect of 
 longitudinal vortex motion driven by the spin current or  spin accumulation injected into the superconductor from the attached ferromagnetic electrodes with polarization $\bm P$. We denote $V_s$ the corresponding spin-dependent external bias. For simplicity we assume that the polarization is aligned with the  spin-splitting filed in the superconductor $\bm P \parallel \bm h$.
To calculate the force acting on vortex from the injected spin current we consider the regime of  temperatures close to the critical one $T_c$. 
In this case we can neglect the superconducting corrections to the density of states. This assumption simplifies  expression for spin-dependent part of the distribution function which can be taken in the form corresponding to the normal metal 
 $f_{T3}(\bm r)=\mu_z (\bm r) \partial_\varepsilon f_0$ where  $\mu_z=\mu_z(\bm r)$ is the spatially-inhomogeneous spin accumulation generated by the external bias $V_s$. Besides that here we consider the regime of small fields $B\ll H_{c2}$ when vortices can be considered as individual objects. 
 The force acting on the single  vortex from non-equilibrium spin-polarized
 environment  $\bm F_d$ can be calculated  using the known general expression  \cite{Larkin1986,KopninBook}.   
 Near the critical temperature when $|\Delta|\ll k_BT$ we obtain the simple analytical result 
 $\bm F_d \approx -\nabla\mu_z S_{loc} |\Delta|^2/(k_BT_c)^2 $,
 where $S_{loc}$ is the total spin localized in the vortex core.
 This driving force, balanced by the friction $\bm F_v=-\rho \bm v_L$, where $\rho$ is the vortex viscosity coefficient,
 yields the flux-flow velocity $\bm v_L\parallel \nabla\mu_z$. 
Its absolute value can be found using the known analytical expression for viscosity coefficient 
 $\rho = \phi_0\sigma_{n}\beta H_{c2}/c^2$. The temperature dependence close to $T_c$ is determined by the coefficient $\beta =\beta_0/ \sqrt{1-T/T_c}$, where $\beta_0$ is some numerical value\cite{osti_7210754, LarkinOvchinnikovTc2, PhysRevB.96.214507}.  Taking into account that the concentration of 
 vortices is determined by the average  magnetic induction $B$ and using the usual expression for the sample-average electric field  
 $\bm E= - \bm v_L\times \bm B/c$ we obtain the  relation
\begin{equation} \label{Eq:Eestimation}
\frac{eE}{\nabla \mu_s} \approx \frac{h}{q}\frac{|\Delta|^2}{(k_BT_c)^2} \frac{B}{H_{c2}}\sqrt{1-\frac{T}{T_c}}.
 \end{equation}
 The obtained result (\ref{Eq:Eestimation}) yields
 the linear response relation for the inverse spin Hall effect because the electric field $E$ and the corresponding electric current are generated in response to the applied spin-dependent voltage $V_s$.
 The overall  temperature dependence of the generated electric field 
$E\propto (1-T/T_c)^{3/2}$ 
 is determined by the order parameter amplitude $\Delta^2 \propto (1-T/T_c)$ and the additional factor which comes from the divergence of vortex viscosity coefficient close to the critical temperature\cite{Larkin1977}  
 $\rho\propto 1/\sqrt{1-T/T_c}$. 

}
\section*{Discussion and conclusions}
 
 We have found the spin current generation by moving vortices which penetrate the whole volume of the type-II superconductor. Thus the obtained spin current in contrast to the previously known schemes based on the injection mechanisms exists everywhere in the sample volume and is prone to the spin relaxation mechanisms such as the spin-flip scattering.   
 The predicted spin current generation can be tested in the open circuit geometries when the vortices annihilate at the insulating boundary. 
 In this case the net spin current at the boundary, $y=0$, should vanish $j_{sy} (y=0) =0$ generating the surface 
  spin accumulation $V_s = \mu_z (y=0)/e$  which can be measured by the ferromagnetic detector electrodes 
  \cite{Huebler2012,Wolf2014,Quay2013,Quay2016}. 
  In the regime when spin relaxation length $l_s$
 is much larger than the intervortex distance, the time-average spin accumulation at low temperatures reads as
 $V_s = -l_s\theta_{sH} Ee^{-y/l_s}$.

The second possible experimental test is based on the direct measurement of the spin current 
   injected through superconductor interfaces into the inverse spin Hall detector\cite{Chumak2012,Hahn2013}. 
   This approach allows to measure both the DC and the high-frequency AC spin current signals. The latter is generated due to the periodic structure of moving vortex lattice. 
   The distribution of the space-periodic spin current component is shown in Fig.\ref{f2b}C,D. The amplitude of AC component  flowing through the superconductor interface, $\langle \tilde j_{sy}\rangle_x$,
     is determined by the variations of the  current average along the boundary,
   $\langle j_{sy}\rangle_x$, with respect to the constant background current $\langle j_{s}\rangle$. 
  At low temperatures, the relative magnitude is given by  
   $\langle \tilde j_{sy}\rangle_x/\langle j_{s}\rangle = 
  (1- \langle\Delta^2\rangle_x/\langle\Delta^2\rangle )(1+\alpha^2)/2$. 
According to the recent measurements, the frequency of vortex entry into the superconducting sample 
   can reach dozens of gigahertz in Pb \cite{Embon2017} and the teraherz range in layered high-temperature
   superconductors \cite{Welp2013}.
  In the suggested setup this is the frequency of the AC spin current generated by the vortex motion. 
   The suggested  high-frequency spin current generation can be useful in 
   antiferromagnetic spintronics
   characterized by the terahertz-range dynamics of the magnetic system\cite{Jungwirth2016}. 
   
   Traditionally the charge imbalance  and spin accumulation has been accessed experimentally using non-local conductance measurements \cite{clarke79,
PhysRevLett.28.1363, PhysRevB.12.4909, hubler2010charge,
Quay2013, quay2015-qsr, Quay2016, Huebler2012,
Kolenda2016, Kolenda2017, Wolf2014}, when the non-equilibrium states were created by the current in the injector circuit. The non-local electric signal has been measured between the normal detector electrodes, either ferromagnetic or non-ferromagnetic attached to the different points of superconducting sample.
Here we show that in the flux-flow regime  the non-equilibrium states  with non-zero charge imbalance $\mu$  and spin accumulation  $\mu_z$ appear in the absence of quasiparticle injector current, but rather just due to the vortex motion. The quantities $\mu$ and $\mu_z$ can be measured using the same electrical detection circuits as in the non-local 
conductance measurement setups. For example,  the tunneling current at the non-ferromagnetic normal detector electrode 
is proportional to $\mu$. In case of the ferromagnetic electrode there is a contribution to the detector current\cite{SilaevRMP2017} proportional to $\mu_z$.
In the flux-flow regime each vortex carries the  distributions of $\mu$ and $\mu_z$ localized in the vortex core. Thus, moving vortices passing close to the detector electrode are expected to generate pulses of the tunneling current  or voltage, depending on the detection scheme. This provides a tool capable for detecting the motion of individual vortices. In contrast to the magnetometer techniques  it does not have the frequency limitations\cite{Embon2017} and therefore can directly resolve the ultrafast vortex motion with the frequencies up to the dozens of gigahertz .

   To conclude, we have demonstrated fundamental mechanisms of direct and inverse spin Hall effects due to 
   the flux-flow of Abrikosov vortices in type-II superconductors. 
   The pure spin current carried by the fast vortices moving in the transverse direction is characterized 
   by the large spin Hall angle which 
   in general does not contain any  small parameters. Besides that
    there is also an AC component which appears due to the periodic structure
     of the vortex lattice. The AC spin current has the same order of magnitude as the average one. 
     This effect can be used for the generation of spin signals in wide frequency domain up to the range of therahertz. 
  We pointed out the longitudinal driving force exerted on  vortex by the injected spin current.
     The vortex motion generated by this force leads to the  inverse spin Hall effect. This mechanism can be applied for flux-flow based detection of pure spin currents.

\section*{Acknowledgements}
   This work was supported by the Academy of Finland. It is our pleasure to acknowledge discussions with 
   Jan Aarts, Tero T. Heikkil\"{a} and Alexander Mel'nikov.

\section*{Author contributions}

Both authors, AV and MS contributed equally to  calculations and writing of the manuscript.

\section*{Additional information}

{\bf Supplementary information} accompanies this paper.

{\bf Competing Interests:} The authors declare that they have no competing interests.

{\bf Data availability:} No datasets were generated or analysed during the current study.

\bibliography{spinhall}

\begin{thebibliography}{10}
\urlstyle{rm}
\expandafter\ifx\csname url\endcsname\relax
  \def\url#1{\texttt{#1}}\fi
\expandafter\ifx\csname urlprefix\endcsname\relax\def\urlprefix{URL }\fi
\expandafter\ifx\csname doiprefix\endcsname\relax\def\doiprefix{DOI: }\fi
\providecommand{\bibinfo}[2]{#2}
\providecommand{\eprint}[2][]{\url{#2}}

\bibitem{Sinova2015}
\bibinfo{author}{Sinova, J.}, \bibinfo{author}{Valenzuela, S.~O.},
  \bibinfo{author}{Wunderlich, J.}, \bibinfo{author}{Back, C.~H.} \&
  \bibinfo{author}{Jungwirth, T.}
\newblock \bibinfo{journal}{\bibinfo{title}{Spin hall effects}}.
\newblock {\emph{\JournalTitle{Rev. Mod. Phys.}}}
  \textbf{\bibinfo{volume}{87}}, \bibinfo{pages}{1213--1260}
  (\bibinfo{year}{2015}).

\bibitem{Abanin2011}
\bibinfo{author}{Abanin, D.~A.} \emph{et~al.}
\newblock \bibinfo{journal}{\bibinfo{title}{Giant nonlocality near the dirac
  point in graphene}}.
\newblock {\emph{\JournalTitle{Science}}} \textbf{\bibinfo{volume}{332}},
  \bibinfo{pages}{328} (\bibinfo{year}{2011}).

\bibitem{Abanin2011a}
\bibinfo{author}{Abanin, D.~A.}, \bibinfo{author}{Gorbachev, R.~V.},
  \bibinfo{author}{Novoselov, K.~S.}, \bibinfo{author}{Geim, A.~K.} \&
  \bibinfo{author}{Levitov, L.~S.}
\newblock \bibinfo{journal}{\bibinfo{title}{Giant spin-hall effect induced by
  the zeeman interaction in graphene}}.
\newblock {\emph{\JournalTitle{Phys. Rev. Lett.}}}
  \textbf{\bibinfo{volume}{107}}, \bibinfo{pages}{096601}
  (\bibinfo{year}{2011}).

\bibitem{Wei2016}
\bibinfo{author}{Wei, P.} \emph{et~al.}
\newblock \bibinfo{journal}{\bibinfo{title}{Strong interfacial exchange field
  in the graphene/eus heterostructure}}.
\newblock {\emph{\JournalTitle{Nature Materials}}}
  \textbf{\bibinfo{volume}{15}}, \bibinfo{pages}{711} (\bibinfo{year}{2016}).

\bibitem{SilaevRMP2017}
\bibinfo{author}{Bergeret, F.}, \bibinfo{author}{Silaev, M.},
  \bibinfo{author}{Virtanen, P.} \& \bibinfo{author}{Heikkila, T.}
\newblock \bibinfo{journal}{\bibinfo{title}{Nonequilibrium effects in
  superconductors with a spin-splitting field}}.
\newblock {\emph{\JournalTitle{arXiv:1706.08245}}}  (\bibinfo{year}{2017}).
\newblock \bibinfo{note}{Accepted to Rev. Mod. Phys.}

\bibitem{Tedrow1971}
\bibinfo{author}{Tedrow, P.~M.} \& \bibinfo{author}{Meservey, R.}
\newblock \bibinfo{journal}{\bibinfo{title}{Spin-dependent tunneling into
  ferromagnetic nickel}}.
\newblock {\emph{\JournalTitle{Phys. Rev. Lett.}}}
  \textbf{\bibinfo{volume}{26}}, \bibinfo{pages}{192--195}
  (\bibinfo{year}{1971}).

\bibitem{Silaev2015}
\bibinfo{author}{Silaev, M.}, \bibinfo{author}{Virtanen, P.},
  \bibinfo{author}{Bergeret, F.~S.} \& \bibinfo{author}{Heikkil\"a, T.~T.}
\newblock \bibinfo{journal}{\bibinfo{title}{Long-range spin accumulation from
  heat injection in mesoscopic superconductors with zeeman splitting}}.
\newblock {\emph{\JournalTitle{Phys. Rev. Lett.}}}
  \textbf{\bibinfo{volume}{114}}, \bibinfo{pages}{167002}
  (\bibinfo{year}{2015}).

\bibitem{Bobkova2015}
\bibinfo{author}{Bobkova, I.~V.} \& \bibinfo{author}{Bobkov, A.~M.}
\newblock \bibinfo{journal}{\bibinfo{title}{Long-range spin imbalance in
  mesoscopic superconductors under zeeman splitting}}.
\newblock {\emph{\JournalTitle{JETP Letters}}} \textbf{\bibinfo{volume}{101}},
  \bibinfo{pages}{118--124} (\bibinfo{year}{2015}).

\bibitem{Krishtop2015}
\bibinfo{author}{Krishtop, T.}, \bibinfo{author}{Houzet, M.} \&
  \bibinfo{author}{Meyer, J.~S.}
\newblock \bibinfo{journal}{\bibinfo{title}{Nonequilibrium spin transport in
  zeeman-split superconductors}}.
\newblock {\emph{\JournalTitle{Phys. Rev. B}}} \textbf{\bibinfo{volume}{91}},
  \bibinfo{pages}{121407} (\bibinfo{year}{2015}).

\bibitem{Virtanen2016}
\bibinfo{author}{Virtanen, P.}, \bibinfo{author}{Heikkilä, T.~T.} \&
  \bibinfo{author}{Bergeret, F.~S.}
\newblock \bibinfo{journal}{\bibinfo{title}{Stimulated quasiparticles in
  spin-split superconductors}}.
\newblock {\emph{\JournalTitle{Phys. Rev. B}}} \textbf{\bibinfo{volume}{93}},
  \bibinfo{pages}{014512} (\bibinfo{year}{2016}).

\bibitem{PhysRevB.98.024516}
\bibinfo{author}{Aikebaier, F.}, \bibinfo{author}{Silaev, M.~A.} \&
  \bibinfo{author}{Heikkil\"a, T.~T.}
\newblock \bibinfo{journal}{\bibinfo{title}{Supercurrent-induced charge-spin
  conversion in spin-split superconductors}}.
\newblock {\emph{\JournalTitle{Phys. Rev. B}}} \textbf{\bibinfo{volume}{98}},
  \bibinfo{pages}{024516}, \doiprefix\url{10.1103/PhysRevB.98.024516}
  (\bibinfo{year}{2018}).

\bibitem{1706.08245}
\bibinfo{author}{Bergeret, F.~S.}, \bibinfo{author}{Silaev, M.},
  \bibinfo{author}{Virtanen, P.} \& \bibinfo{author}{Heikkil\"a, T.~T.}
\newblock \bibinfo{title}{Nonequilibrium effects in superconductors with a
  spin-splitting field} (\bibinfo{year}{2017}).
\newblock \eprint{arXiv:1706.08245}.

\bibitem{Huebler2012}
\bibinfo{author}{H\"ubler, F.}, \bibinfo{author}{Wolf, M.~J.},
  \bibinfo{author}{Beckmann, D.} \& \bibinfo{author}{v.~L\"ohneysen, H.}
\newblock \bibinfo{journal}{\bibinfo{title}{Long-range spin-polarized
  quasiparticle transport in mesoscopic al superconductors with a zeeman
  splitting}}.
\newblock {\emph{\JournalTitle{Phys. Rev. Lett.}}}
  \textbf{\bibinfo{volume}{109}}, \bibinfo{pages}{207001}
  (\bibinfo{year}{2012}).

\bibitem{Wolf2014}
\bibinfo{author}{Wolf, M.~J.}, \bibinfo{author}{S\"urgers, C.},
  \bibinfo{author}{Fischer, G.} \& \bibinfo{author}{Beckmann, D.}
\newblock \bibinfo{journal}{\bibinfo{title}{Spin-polarized quasiparticle
  transport in exchange-split superconducting aluminum on europium sulfide}}.
\newblock {\emph{\JournalTitle{Phys. Rev. B}}} \textbf{\bibinfo{volume}{90}},
  \bibinfo{pages}{144509} (\bibinfo{year}{2014}).

\bibitem{Quay2013}
\bibinfo{author}{Quay, C. H.~L.}, \bibinfo{author}{Chevallier, D.},
  \bibinfo{author}{Bena, C.} \& \bibinfo{author}{Aprili, M.}
\newblock \bibinfo{journal}{\bibinfo{title}{Spin imbalance and spin-charge
  separation in a mesoscopic superconductor}}.
\newblock {\emph{\JournalTitle{Nat Phys}}} \textbf{\bibinfo{volume}{9}},
  \bibinfo{pages}{84--88} (\bibinfo{year}{2013}).

\bibitem{Quay2016}
\bibinfo{author}{Quay, C. H.~L.}, \bibinfo{author}{Dutreix, C.},
  \bibinfo{author}{Chevallier, D.}, \bibinfo{author}{Bena, C.} \&
  \bibinfo{author}{Aprili, M.}
\newblock \bibinfo{journal}{\bibinfo{title}{Frequency-domain measurement of the
  spin-imbalance lifetime in superconductors}}.
\newblock {\emph{\JournalTitle{Phys. Rev. B}}} \textbf{\bibinfo{volume}{93}},
  \bibinfo{pages}{220501} (\bibinfo{year}{2016}).

\bibitem{Maki1966}
\bibinfo{author}{Maki, K.}
\newblock \bibinfo{journal}{\bibinfo{title}{Effect of pauli paramagnetism on
  magnetic properties of high-field superconductors}}.
\newblock {\emph{\JournalTitle{Phys. Rev.}}} \textbf{\bibinfo{volume}{148}},
  \bibinfo{pages}{362--369} (\bibinfo{year}{1966}).

\bibitem{Chandrasekhar1962}
\bibinfo{author}{Chandrasekhar, B.~S.}
\newblock \bibinfo{journal}{\bibinfo{title}{A note on the maximum critical
  field of high‐field superconductors}}.
\newblock {\emph{\JournalTitle{Appl. Phys. Lett.}}}
  \textbf{\bibinfo{volume}{1}}, \bibinfo{pages}{7--8},
  \doiprefix\url{10.1063/1.1777362} (\bibinfo{year}{1962}).

\bibitem{Clogston1962}
\bibinfo{author}{Clogston, A.~M.}
\newblock \bibinfo{journal}{\bibinfo{title}{Upper limit for the critical field
  in hard superconductors}}.
\newblock {\emph{\JournalTitle{Phys. Rev. Lett.}}}
  \textbf{\bibinfo{volume}{9}}, \bibinfo{pages}{266--267}
  (\bibinfo{year}{1962}).

\bibitem{saint1969type}
\bibinfo{author}{Saint-James, D.}, \bibinfo{author}{Sarma, G.} \&
  \bibinfo{author}{Thomas, E.}
\newblock \emph{\bibinfo{title}{Type II Superconductivity}}.
\newblock Commonwealth and International Library. Liberal Studies Divi
  (\bibinfo{publisher}{Elsevier Science \& Technology}, \bibinfo{year}{1969}).

\bibitem{Gurevich2010}
\bibinfo{author}{Gurevich, A.}
\newblock \bibinfo{journal}{\bibinfo{title}{Upper critical field and the
  fulde-ferrel-larkin-ovchinnikov transition in multiband superconductors}}.
\newblock {\emph{\JournalTitle{Phys. Rev. B}}} \textbf{\bibinfo{volume}{82}},
  \bibinfo{pages}{184504} (\bibinfo{year}{2010}).

\bibitem{Cho2011}
\bibinfo{author}{Cho, K.} \emph{et~al.}
\newblock \bibinfo{journal}{\bibinfo{title}{Anisotropic upper critical field
  and possible fulde-ferrel-larkin-ovchinnikov state in the stoichiometric
  pnictide superconductor lifeas}}.
\newblock {\emph{\JournalTitle{Phys. Rev. B}}} \textbf{\bibinfo{volume}{83}},
  \bibinfo{pages}{060502} (\bibinfo{year}{2011}).

\bibitem{PhysRevLett.24.1004}
\bibinfo{author}{Tedrow, P.~M.}, \bibinfo{author}{Meservey, R.} \&
  \bibinfo{author}{Schwartz, B.~B.}
\newblock \bibinfo{journal}{\bibinfo{title}{Experimental evidence for a
  first-order magnetic transition in thin superconducting aluminum films}}.
\newblock {\emph{\JournalTitle{Phys. Rev. Lett.}}}
  \textbf{\bibinfo{volume}{24}}, \bibinfo{pages}{1004--1007},
  \doiprefix\url{10.1103/PhysRevLett.24.1004} (\bibinfo{year}{1970}).

\bibitem{Kolenda2016}
\bibinfo{author}{Kolenda, S.}, \bibinfo{author}{Wolf, M.~J.} \&
  \bibinfo{author}{Beckmann, D.}
\newblock \bibinfo{journal}{\bibinfo{title}{Observation of thermoelectric
  currents in high-field superconductor-ferromagnet tunnel junctions}}.
\newblock {\emph{\JournalTitle{Phys. Rev. Lett.}}}
  \textbf{\bibinfo{volume}{116}}, \bibinfo{pages}{097001}
  (\bibinfo{year}{2016}).

\bibitem{fulde_ferrell.1964}
\bibinfo{author}{Fulde, P.} \& \bibinfo{author}{Ferrell, R.~A.}
\newblock \bibinfo{journal}{\bibinfo{title}{Superconductivity in a strong
  spin-exchange field}}.
\newblock {\emph{\JournalTitle{Phys. Rev.}}} \textbf{\bibinfo{volume}{135}},
  \bibinfo{pages}{A550--A563}, \doiprefix\url{10.1103/PhysRev.135.A550}
  (\bibinfo{year}{1964}).

\bibitem{Kolenda2017}
\bibinfo{author}{Kolenda, S.}, \bibinfo{author}{S\"urgers, C.},
  \bibinfo{author}{Fischer, G.} \& \bibinfo{author}{Beckmann, D.}
\newblock \bibinfo{journal}{\bibinfo{title}{Thermoelectric effects in
  superconductor-ferromagnet tunnel junctions on europium sulfide}}.
\newblock {\emph{\JournalTitle{Phys. Rev. B}}} \textbf{\bibinfo{volume}{95}},
  \bibinfo{pages}{224505} (\bibinfo{year}{2017}).

\bibitem{Strambini2017}
\bibinfo{author}{Strambini, E.} \emph{et~al.}
\newblock \bibinfo{journal}{\bibinfo{title}{Revealing the magnetic proximity
  effect in eus/al bilayers through superconducting tunneling spectroscopy}}.
\newblock {\emph{\JournalTitle{Phys. Rev. Materials}}}
  \textbf{\bibinfo{volume}{1}}, \bibinfo{pages}{054402} (\bibinfo{year}{2017}).

\bibitem{PhysRevB.97.224414}
\bibinfo{author}{Yao, Y.} \emph{et~al.}
\newblock \bibinfo{journal}{\bibinfo{title}{Probe of spin dynamics in
  superconducting nbn thin films via spin pumping}}.
\newblock {\emph{\JournalTitle{Phys. Rev. B}}} \textbf{\bibinfo{volume}{97}},
  \bibinfo{pages}{224414}, \doiprefix\url{10.1103/PhysRevB.97.224414}
  (\bibinfo{year}{2018}).

\bibitem{PhysRevB.38.8823}
\bibinfo{author}{Tokuyasu, T.}, \bibinfo{author}{Sauls, J.~A.} \&
  \bibinfo{author}{Rainer, D.}
\newblock \bibinfo{journal}{\bibinfo{title}{Proximity effect of a ferromagnetic
  insulator in contact with a superconductor}}.
\newblock {\emph{\JournalTitle{Phys. Rev. B}}} \textbf{\bibinfo{volume}{38}},
  \bibinfo{pages}{8823--8833}, \doiprefix\url{10.1103/PhysRevB.38.8823}
  (\bibinfo{year}{1988}).

\bibitem{1709.08856}
\bibinfo{author}{Heikkil\"a, T.~T.}, \bibinfo{author}{Ojaj\"arvi, R.},
  \bibinfo{author}{Maasilta, I.~J.}, \bibinfo{author}{Giazotto, F.} \&
  \bibinfo{author}{Bergeret, F.}
\newblock \bibinfo{title}{Thermoelectric radiation detector based on
  superconductor/ferromagnet systems} (\bibinfo{year}{2017}).
\newblock \eprint{arXiv:1709.08856}.

\bibitem{PhysRevB.98.020501}
\bibinfo{author}{Virtanen, P.}, \bibinfo{author}{Bergeret, F.~S.},
  \bibinfo{author}{Strambini, E.}, \bibinfo{author}{Giazotto, F.} \&
  \bibinfo{author}{Braggio, A.}
\newblock \bibinfo{journal}{\bibinfo{title}{Majorana bound states in hybrid
  two-dimensional josephson junctions with ferromagnetic insulators}}.
\newblock {\emph{\JournalTitle{Phys. Rev. B}}} \textbf{\bibinfo{volume}{98}},
  \bibinfo{pages}{020501}, \doiprefix\url{10.1103/PhysRevB.98.020501}
  (\bibinfo{year}{2018}).

\bibitem{Chandrasekhar:1962}
\bibinfo{author}{Chandrasekhar, B.~S.}
\newblock \bibinfo{journal}{\bibinfo{title}{A note on the maximum critical
  field of high-field superconductors}}.
\newblock {\emph{\JournalTitle{Appl. Phys. Lett.}}}
  \textbf{\bibinfo{volume}{1}}, \bibinfo{pages}{7--8},
  \doiprefix\url{10.1063/1.1777362} (\bibinfo{year}{1962}).

\bibitem{PhysRevLett.9.266}
\bibinfo{author}{Clogston, A.~M.}
\newblock \bibinfo{journal}{\bibinfo{title}{Upper limit for the critical field
  in hard superconductors}}.
\newblock {\emph{\JournalTitle{Phys. Rev. Lett.}}}
  \textbf{\bibinfo{volume}{9}}, \bibinfo{pages}{266--267},
  \doiprefix\url{10.1103/PhysRevLett.9.266} (\bibinfo{year}{1962}).

\bibitem{larkin_ovchinnikov.1964}
\bibinfo{author}{Larkin, A.~I.} \& \bibinfo{author}{Ovchinnikov, Y.~N.}
\newblock \bibinfo{journal}{\bibinfo{title}{Inhomogeneous state of
  superconductors}}.
\newblock {\emph{\JournalTitle{Sov. Phys. JETP}}}
  \textbf{\bibinfo{volume}{20}}, \bibinfo{pages}{762--769}
  (\bibinfo{year}{1965}).

\bibitem{Aslamazov}
\bibinfo{author}{Aslamazov, L.}
\newblock \bibinfo{journal}{\bibinfo{title}{Influence of impurities on the
  existence of an inhomogeneous state in a ferromagnetic superconductor}}.
\newblock {\emph{\JournalTitle{Journal of Experimental and Theoretical
  Physics}}} \textbf{\bibinfo{volume}{28}}, \bibinfo{pages}{773}
  (\bibinfo{year}{1969}).

\bibitem{Buzdin2005}
\bibinfo{author}{Buzdin, A.~I.}
\newblock \bibinfo{journal}{\bibinfo{title}{Proximity effects in
  superconductor-ferromagnet heterostructures}}.
\newblock {\emph{\JournalTitle{Rev. Mod. Phys.}}}
  \textbf{\bibinfo{volume}{77}}, \bibinfo{pages}{935--976}
  (\bibinfo{year}{2005}).

\bibitem{1710.10833}
\bibinfo{author}{Yao, Y.} \emph{et~al.}
\newblock \bibinfo{title}{Probe of spin dynamics in superconducting nbn thin
  films via spin pumping} (\bibinfo{year}{2017}).
\newblock \eprint{arXiv:1710.10833}.

\bibitem{Buzdin1984}
\bibinfo{author}{Buzdin, A.~I.}, \bibinfo{author}{Bulaevskiĭ, L.~N.},
  \bibinfo{author}{Kulich, M.~L.} \& \bibinfo{author}{Panyukov, S.~V.}
\newblock \bibinfo{journal}{\bibinfo{title}{Magnetic superconductors}}.
\newblock {\emph{\JournalTitle{Soviet Physics Uspekhi}}}
  \textbf{\bibinfo{volume}{27}}, \bibinfo{pages}{927} (\bibinfo{year}{1984}).

\bibitem{Mueller2001}
\bibinfo{author}{Müller, K.-H.} \& \bibinfo{author}{Narozhnyi, V.~N.}
\newblock \bibinfo{journal}{\bibinfo{title}{Interaction of superconductivity
  and magnetism in borocarbide superconductors}}.
\newblock {\emph{\JournalTitle{Reports on Progress in Physics}}}
  \textbf{\bibinfo{volume}{64}}, \bibinfo{pages}{943} (\bibinfo{year}{2001}).

\bibitem{Bluhm2006}
\bibinfo{author}{Bluhm, H.}, \bibinfo{author}{Sebastian, S.~E.},
  \bibinfo{author}{Guikema, J.~W.}, \bibinfo{author}{Fisher, I.~R.} \&
  \bibinfo{author}{Moler, K.~A.}
\newblock \bibinfo{journal}{\bibinfo{title}{Scanning hall probe imaging of
  ${\mathrm{erni}}_{2}{\mathrm{b}}_{2}\mathrm{C}$}}.
\newblock {\emph{\JournalTitle{Phys. Rev. B}}} \textbf{\bibinfo{volume}{73}},
  \bibinfo{pages}{014514} (\bibinfo{year}{2006}).

\bibitem{DeBeer-Schmitt2007}
\bibinfo{author}{DeBeer-Schmitt, L.} \emph{et~al.}
\newblock \bibinfo{journal}{\bibinfo{title}{Pauli paramagnetic effects on
  vortices in superconducting
  ${\mathrm{tmni}}_{2}{\mathrm{b}}_{2}\mathrm{C}$}}.
\newblock {\emph{\JournalTitle{Phys. Rev. Lett.}}}
  \textbf{\bibinfo{volume}{99}}, \bibinfo{pages}{167001}
  (\bibinfo{year}{2007}).

\bibitem{Schmid1975}
\bibinfo{author}{Schmid, A.} \& \bibinfo{author}{Sch\"on, G.}
\newblock \bibinfo{journal}{\bibinfo{title}{Linearized kinetic equations and
  relaxation processes of a superconductor near t c}}.
\newblock {\emph{\JournalTitle{Journal of Low Temperature Physics}}}
  \textbf{\bibinfo{volume}{20}}, \bibinfo{pages}{207--227}
  (\bibinfo{year}{1975}).

\bibitem{Belzig1999}
\bibinfo{author}{Belzig, W.}, \bibinfo{author}{Wilhelm, F.~K.},
  \bibinfo{author}{Bruder, C.}, \bibinfo{author}{Schön, G.} \&
  \bibinfo{author}{Zaikin, A.~D.}
\newblock \bibinfo{journal}{\bibinfo{title}{Quasiclassical green’s function
  approach to mesoscopic superconductivity}}.
\newblock {\emph{\JournalTitle{Superlattices and Microstructures}}}
  \textbf{\bibinfo{volume}{25}}, \bibinfo{pages}{1251--1288}
  (\bibinfo{year}{1999}).

\bibitem{Embon2017}
\bibinfo{author}{Embon, L.} \emph{et~al.}
\newblock \bibinfo{journal}{\bibinfo{title}{Imaging of super-fast dynamics and
  flow instabilities of superconducting vortices}}.
\newblock {\emph{\JournalTitle{Nature Communications}}}
  \textbf{\bibinfo{volume}{8}}, \bibinfo{pages}{85} (\bibinfo{year}{2017}).

\bibitem{nonlin1}
\bibinfo{journal}{\bibinfo{author}{Larkin, A.~I.} \&
  \bibinfo{author}{Ovchinnikov, Y.~N.}}
\newblock {\emph{\JournalTitle{Sov. Phys. JETP}}}
  \textbf{\bibinfo{volume}{41}}, \bibinfo{pages}{960} (\bibinfo{year}{1976}).

\bibitem{nonlin2}
\bibinfo{journal}{\bibinfo{author}{Klein, W.}, \bibinfo{author}{Huebener,
  R.~P.}, \bibinfo{author}{Gauss, S.} \& \bibinfo{author}{Parisi, J.}}
\newblock {\emph{\JournalTitle{J. Low Temp. Phys.}}}
  \textbf{\bibinfo{volume}{61}}, \bibinfo{pages}{413} (\bibinfo{year}{1985}).

\bibitem{Larkin1986}
\bibinfo{author}{Larkin, A.~I.} \& \bibinfo{author}{Ovchinnikov, Y.~N.}
\newblock In \bibinfo{editor}{Langenberg, D.~N.} \& \bibinfo{editor}{Larkin,
  A.} (eds.) \emph{\bibinfo{booktitle}{Modern Problems in Condensed Matter
  Sciences: Nonequilibrium Superconductivity}}, \bibinfo{pages}{493}
  (\bibinfo{publisher}{Elsevier}, \bibinfo{year}{1986}).

\bibitem{KopninBook}
\bibinfo{author}{Kopnin, N.~B.}
\newblock \emph{\bibinfo{title}{Theory of Nonequilibrium Superconductivity}}
  (\bibinfo{publisher}{Oxford University Press}, \bibinfo{year}{2001}).

\bibitem{PhysRev.164.591}
\bibinfo{author}{Caroli, C.} \& \bibinfo{author}{Maki, K.}
\newblock \bibinfo{journal}{\bibinfo{title}{Motion of the vortex structure in
  type-ii superconductors in high magnetic field}}.
\newblock {\emph{\JournalTitle{Phys. Rev.}}} \textbf{\bibinfo{volume}{164}},
  \bibinfo{pages}{591--607}, \doiprefix\url{10.1103/PhysRev.164.591}
  (\bibinfo{year}{1967}).

\bibitem{Silaev2016a}
\bibinfo{author}{Silaev, M.} \& \bibinfo{author}{Vargunin, A.}
\newblock \bibinfo{journal}{\bibinfo{title}{Vortex motion and flux-flow
  resistivity in dirty multiband superconductors}}.
\newblock {\emph{\JournalTitle{Phys. Rev. B}}} \textbf{\bibinfo{volume}{94}},
  \bibinfo{pages}{224506} (\bibinfo{year}{2016}).

\bibitem{PhysRevB.1.327}
\bibinfo{author}{Thompson, R.~S.}
\newblock \bibinfo{journal}{\bibinfo{title}{Microwave, flux flow, and
  fluctuation resistance of dirty type-ii superconductors}}.
\newblock {\emph{\JournalTitle{Phys. Rev. B}}} \textbf{\bibinfo{volume}{1}},
  \bibinfo{pages}{327--333}, \doiprefix\url{10.1103/PhysRevB.1.327}
  (\bibinfo{year}{1970}).

\bibitem{Caroli1966}
\bibinfo{author}{Caroli, C.}, \bibinfo{author}{Cyrot, M.} \&
  \bibinfo{author}{de~Gennes, P.~G.}
\newblock \bibinfo{journal}{\bibinfo{title}{The magnetic behavior of dirty
  superconductors}}.
\newblock {\emph{\JournalTitle{Solid State Communications}}}
  \textbf{\bibinfo{volume}{4}}, \bibinfo{pages}{17--19} (\bibinfo{year}{1966}).

\bibitem{Silaev2016}
\bibinfo{author}{Silaev, M.}
\newblock \bibinfo{journal}{\bibinfo{title}{Magnetic behavior of dirty
  multiband superconductors near the upper critical field}}.
\newblock {\emph{\JournalTitle{Phys. Rev. B}}} \textbf{\bibinfo{volume}{93}},
  \bibinfo{pages}{214509} (\bibinfo{year}{2016}).

\bibitem{Kleiner1964}
\bibinfo{author}{Kleiner, W.~H.}, \bibinfo{author}{Roth, L.~M.} \&
  \bibinfo{author}{Autler, S.~H.}
\newblock \bibinfo{journal}{\bibinfo{title}{Bulk solution of ginzburg-landau
  equations for type ii superconductors: Upper critical field region}}.
\newblock {\emph{\JournalTitle{Phys. Rev.}}} \textbf{\bibinfo{volume}{133}},
  \bibinfo{pages}{A1226--A1227} (\bibinfo{year}{1964}).

\bibitem{Maki1969}
\bibinfo{journal}{\bibinfo{author}{Maki, K.}}
\newblock {\emph{\JournalTitle{J. Low Temp. Phys.}}}
  \textbf{\bibinfo{volume}{1}}, \bibinfo{pages}{45} (\bibinfo{year}{1969}).

\bibitem{Hunter2007}
\bibinfo{author}{Hunter, J.~D.}
\newblock \bibinfo{journal}{\bibinfo{title}{Matplotlib: A 2d graphics
  environment}}.
\newblock {\emph{\JournalTitle{Computing In Science \& Engineering}}}
  \textbf{\bibinfo{volume}{9}}, \bibinfo{pages}{IEEE COMPUTER SOC---95}
  (\bibinfo{year}{2007}).

\bibitem{yagi2006charge}
\bibinfo{author}{Yagi, R.}
\newblock \bibinfo{journal}{\bibinfo{title}{Charge imbalance observed in
  voltage-biased superconductor--normal tunnel junctions}}.
\newblock {\emph{\JournalTitle{Phys. Rev. B}}} \textbf{\bibinfo{volume}{73}},
  \bibinfo{pages}{134507}, \doiprefix\url{10.1103/PhysRevB.73.134507}
  (\bibinfo{year}{2006}).

\bibitem{hubler2010charge}
\bibinfo{author}{H{\"u}bler, F.}, \bibinfo{author}{Lemyre, J.~C.},
  \bibinfo{author}{Beckmann, D.} \& \bibinfo{author}{v.~L{\"o}hneysen, H.}
\newblock \bibinfo{journal}{\bibinfo{title}{Charge imbalance in superconductors
  in the low-temperature limit}}.
\newblock {\emph{\JournalTitle{Phys. Rev. B}}} \textbf{\bibinfo{volume}{81}},
  \bibinfo{pages}{184524}, \doiprefix\url{10.1103/PhysRevB.81.184524}
  (\bibinfo{year}{2010}).

\bibitem{Tinkham1972}
\bibinfo{author}{Tinkham, M.} \& \bibinfo{author}{Clarke, J.}
\newblock \bibinfo{journal}{\bibinfo{title}{Theory of pair-quasiparticle
  potential difference in nonequilibrium superconductors}}.
\newblock {\emph{\JournalTitle{Phys. Rev. Lett.}}}
  \textbf{\bibinfo{volume}{28}}, \bibinfo{pages}{1366--1369}
  (\bibinfo{year}{1972}).

\bibitem{Tinkham1972a}
\bibinfo{author}{Tinkham, M.}
\newblock \bibinfo{journal}{\bibinfo{title}{Tunneling generation, relaxation,
  and tunneling detection of hole-electron imbalance in superconductors}}.
\newblock {\emph{\JournalTitle{Phys. Rev. B}}} \textbf{\bibinfo{volume}{6}},
  \bibinfo{pages}{1747--1756} (\bibinfo{year}{1972}).

\bibitem{clarke79}
\bibinfo{author}{Clarke, J.}, \bibinfo{author}{Fjordb\o{}ge, B.~R.} \&
  \bibinfo{author}{Lindelof, P.~E.}
\newblock \bibinfo{journal}{\bibinfo{title}{Supercurrent-induced charge
  imbalance measured in a superconductor in the presence of a thermal
  gradient}}.
\newblock {\emph{\JournalTitle{Phys. Rev. Lett.}}}
  \textbf{\bibinfo{volume}{43}}, \bibinfo{pages}{642--645},
  \doiprefix\url{10.1103/PhysRevLett.43.642} (\bibinfo{year}{1979}).

\bibitem{clarke80}
\bibinfo{author}{Clarke, J.} \& \bibinfo{author}{Tinkham, M.}
\newblock \bibinfo{journal}{\bibinfo{title}{Theory of quasiparticle charge
  imbalance induced in a superconductor by a supercurrent in the presence of a
  thermal gradient}}.
\newblock {\emph{\JournalTitle{Phys. Rev. Lett.}}}
  \textbf{\bibinfo{volume}{44}}, \bibinfo{pages}{106--109},
  \doiprefix\url{10.1103/PhysRevLett.44.106} (\bibinfo{year}{1980}).

\bibitem{PhysRevLett.28.1363}
\bibinfo{author}{Clarke, J.}
\newblock \bibinfo{journal}{\bibinfo{title}{Experimental observation of
  pair-quasiparticle potential difference in nonequilibrium superconductors}}.
\newblock {\emph{\JournalTitle{Phys. Rev. Lett.}}}
  \textbf{\bibinfo{volume}{28}}, \bibinfo{pages}{1363--1366},
  \doiprefix\url{10.1103/PhysRevLett.28.1363} (\bibinfo{year}{1972}).

\bibitem{PhysRevB.12.4909}
\bibinfo{author}{Yu, M.~L.} \& \bibinfo{author}{Mercereau, J.~E.}
\newblock \bibinfo{journal}{\bibinfo{title}{Nonequilibrium quasiparticle
  current at superconducting boundaries}}.
\newblock {\emph{\JournalTitle{Phys. Rev. B}}} \textbf{\bibinfo{volume}{12}},
  \bibinfo{pages}{4909--4916}, \doiprefix\url{10.1103/PhysRevB.12.4909}
  (\bibinfo{year}{1975}).

\bibitem{quay2015-qsr}
\bibinfo{author}{Quay, C. H.~L.}, \bibinfo{author}{Chiffaudel, Y.},
  \bibinfo{author}{Strunk, C.} \& \bibinfo{author}{Aprili, M.}
\newblock \bibinfo{journal}{\bibinfo{title}{Quasiparticle spin resonance and
  coherence in superconducting aluminium}}.
\newblock {\emph{\JournalTitle{Nat. Commun.}}} \textbf{\bibinfo{volume}{6}},
  \bibinfo{pages}{8660}, \doiprefix\url{10.1038/ncomms9660}
  (\bibinfo{year}{2015}).

\bibitem{PhysRev.140.A1197}
\bibinfo{author}{Bardeen, J.} \& \bibinfo{author}{Stephen, M.~J.}
\newblock \bibinfo{journal}{\bibinfo{title}{Theory of the motion of vortices in
  superconductors}}.
\newblock {\emph{\JournalTitle{Phys. Rev.}}} \textbf{\bibinfo{volume}{140}},
  \bibinfo{pages}{A1197--A1207}, \doiprefix\url{10.1103/PhysRev.140.A1197}
  (\bibinfo{year}{1965}).

\bibitem{osti_7210754}
\bibinfo{author}{Gor'kov, L.} \& \bibinfo{author}{Kopnin, N.}
\newblock \bibinfo{journal}{\bibinfo{title}{Some features of viscous flow of
  vortices in superconducting alloys near the critical temperature}}.
\newblock {\emph{\JournalTitle{Sov. Phys. - JETP (Engl. Transl.); (United
  States)}}} \textbf{\bibinfo{volume}{37:1}}, \bibinfo{pages}{183}
  (\bibinfo{year}{1973}).

\bibitem{LarkinOvchinnikovTc2}
\bibinfo{journal}{\bibinfo{author}{Larkin, A.} \& \bibinfo{author}{Ovchinnikov,
  Y.}}
\newblock {\emph{\JournalTitle{Sov. Phys. JETP}}}
  \textbf{\bibinfo{volume}{46}}, \bibinfo{pages}{155} (\bibinfo{year}{1977}).

\bibitem{PhysRevB.96.214507}
\bibinfo{author}{Vargunin, A.} \& \bibinfo{author}{Silaev, M.~A.}
\newblock \bibinfo{journal}{\bibinfo{title}{Self-consistent calculation of the
  flux-flow conductivity in diffusive superconductors}}.
\newblock {\emph{\JournalTitle{Phys. Rev. B}}} \textbf{\bibinfo{volume}{96}},
  \bibinfo{pages}{214507}, \doiprefix\url{10.1103/PhysRevB.96.214507}
  (\bibinfo{year}{2017}).

\bibitem{Larkin1977}
\bibinfo{journal}{\bibinfo{author}{Larkin, A.} \& \bibinfo{author}{Ovchinnikov,
  Y.}}
\newblock {\emph{\JournalTitle{Sov. Phys. JETP}}}
  \textbf{\bibinfo{volume}{46}}, \bibinfo{pages}{155} (\bibinfo{year}{1977}).

\bibitem{Chumak2012}
\bibinfo{author}{Chumak, A.~V.} \emph{et~al.}
\newblock \bibinfo{journal}{\bibinfo{title}{Direct detection of magnon spin
  transport by the inverse spin hall effect}}.
\newblock {\emph{\JournalTitle{Appl. Phys. Lett.}}}
  \textbf{\bibinfo{volume}{100}}, \bibinfo{pages}{082405},
  \doiprefix\url{10.1063/1.3689787} (\bibinfo{year}{2012}).

\bibitem{Hahn2013}
\bibinfo{journal}{\bibinfo{author}{Hahn, C.} \emph{et~al.}}
\newblock {\emph{\JournalTitle{Phys. Rev. Lett.}}}
  \textbf{\bibinfo{volume}{111}}, \bibinfo{pages}{217204}
  (\bibinfo{year}{2013}).

\bibitem{Welp2013}
\bibinfo{author}{Welp, U.}, \bibinfo{author}{Kadowaki, K.} \&
  \bibinfo{author}{Kleiner, R.}
\newblock \bibinfo{journal}{\bibinfo{title}{Superconducting emitters of thz
  radiation}}.
\newblock {\emph{\JournalTitle{Nature Photonics}}}
  \textbf{\bibinfo{volume}{7}}, \bibinfo{pages}{702} (\bibinfo{year}{2013}).

\bibitem{Jungwirth2016}
\bibinfo{author}{Jungwirth, T.}, \bibinfo{author}{Marti, X.},
  \bibinfo{author}{Wadley, P.} \& \bibinfo{author}{Wunderlich, J.}
\newblock \bibinfo{journal}{\bibinfo{title}{Antiferromagnetic spintronics}}.
\newblock {\emph{\JournalTitle{Nature Nanotechnology}}}
  \textbf{\bibinfo{volume}{11}}, \bibinfo{pages}{231} (\bibinfo{year}{2016}).

\end{thebibliography}

\end{document}